\documentclass[intlimits,twoside,a4paper]{article}

\usepackage[cp1251]{inputenc}

\usepackage[eqsecnum]{cmpj3}
\issue{2023}{26}{4}{43603}
\doinumber{10.5488/CMP.26.43603}
\title[Thermal transport properties of graphene]%
{Effect of out-of-plane acoustic phonons on the thermal transport properties of graphene}
\author[J. Chen, Y. Liu]{J. Chen\orcid{0000-0002-2710-5641}\thanks{Corresponding author: \email{cjj@hpu.edu.cn}.},
        Y. Liu}
\address{Department of Energy and Power Engineering, School of Mechanical and Power Engineering, Henan Polytechnic University, Jiaozuo, Henan, 454000, P.R. China} 

\date{Received April 19, 2023, in final form July 28, 2023}

\begin{document}

\maketitle

\begin{abstract}
The lattice thermal conductivity of graphene is evaluated using a microscopic model that takes into account the lattice's discrete nature and the phonon dispersion relation within the Brillouin zone. The Boltzmann transport equation is solved iteratively within the framework of three-phonon interactions without taking into account the four-phonon scattering process. The Umklapp and normal collisions are treated rigorously, thereby avoiding relaxation-time and long-wavelength approximations. The mechanisms of the failures of these approximations in predicting the thermal transport properties are discussed. Evaluation of the thermal conductivity is performed at different temperatures and frequencies and in different crystallite sizes. Reasonably good agreement with the experimental data is obtained. The calculation reveals a critical role of out-of-plane acoustic phonons in determining the thermal conductivity. The out-of-plane acoustic phonons contribute greatly and the longitudinal and transverse acoustic phonons make small contributions over a wide range of temperatures and frequencies. The out-of-plane acoustic phonons dominate the thermal conductivity due to their high density of states and restrictions governing the anharmonic phonon scattering. The selection rule severely restricts the phase space for out-of-plane phonon scattering due to reflection symmetry. The optical phonon contribution cannot be neglected at higher temperatures. Both Umklapp and normal processes must be taken into account in order to predict the phonon transport properties accurately.
%
%
%\keywords Up to six keywords (\href{https://physh.aps.org/browse}{Physics Subject Headings})
\printkeywords
%
%\pacs 63.20.-e; 63.20.Kr; 63.20.Ry; 65.80.Ck; 74.25.F-; 81.05.ue
\end{abstract}

\section{Introduction}

%\doclicenseThis

Graphene is a two-dimensional form of carbon that has attracted intense interest on account of its remarkable physical properties \cite{Qui23,Col22}. For example, graphene has a tremendous mechanical strength~\cite{Zan14,Kas19}. Graphene also has extraordinary electrical and thermal transport properties \cite{Zhu09,Bor13}. Graphene offers excellent potential for thermal management applications \cite{Pyu19,Lee15} due to its extremely high thermal conductivity, up to about 5000 W/(m$\cdot$K) \cite{Bal08,Gho08}, and relatively low density. Other carbon-based materials may possess exceedingly similar thermal transport properties, such as single-walled carbon nanotubes~\cite{Pop05,Yu05}, graphite \cite{Sla62,Tyl53}, and diamond \cite{Onn92,Wei93}. The thermal transport in graphene can be primarily attributed to lattice waves \cite{Lin10a,Seo10}, and therefore the thermal conductivity is limited by the scattering of lattice waves from crystallite boundaries \cite{Hel14,Has22}. Heat from electronic circuits may be removed efficiently by using the peculiar thermal transport properties of graphene \cite{Dua22,Yan21}, thereby making it possible to solve thermal management problems at the nanoscale.

In addition to their importance in the acoustic properties, phonons are essential in the phenomenon of heat conduction, for instance, the thermal transport properties of graphene. The vibrations in the graphene lattice can be characterized by the following phonon branches: longitudinal acoustic mode, longitudinal optic mode, transverse acoustic mode, transverse optic mode, out-of-plane acoustic mode, and out-of-plane optical mode. Longitudinal and transverse phonons vibrate in the plane of the layer, while out-of-plane phonons, also referred to as flexural phonons, vibrate out of the plane of the layer. Over a wide range of the Brillouin zone, the longitudinal and transverse acoustic phonons have linear dispersion, whereas the out-of-plane acoustic phonons have quadratic dispersion. The acoustic phonons that vibrate in the plane carry most of the heat, whereas the acoustic phonons that vibrate out of the plane do not significantly contribute to the thermal transport properties \cite{Nik09a,Nik09b}. This is because for a zero wavevector the out-of-plane acoustic phonons have a vanishing group velocity, and there is a relation between strong Umklapp scattering and large out-of-plane mode Gr\"uneisen parameters \cite{Mou05,Bon07}. The effects of longitudinal and transverse acoustic modes are so strong that the out-of-plane acoustic contribution becomes negligible in comparison.

Doubtful or even negative opinions are held on whether out-of-plane acoustic phonons make negligible contribution. Recent theoretical studies have demonstrated that out-of-plane acoustic phonons are of great importance in thermal transport \cite{Pol18,Fen18}. In suspended graphene, the out-of-plane acoustic phonons may carry primarily heat \cite{Seo10}. Additionally, the theoretical predictions are in agreement with the experimental measurements for graphene supported on a silicon dioxide substrate \cite{Seo10}. Consequently, knowledge of lattice dynamics of out-of-plane acoustic phonons may be of vital importance in understanding the thermal properties of graphene \cite{Kua20,Lin18}. However, these studies do not permit entirely satisfactory theoretical interpretation. Refinement of the studies and their extension to a larger variety of graphene-based materials and to a wider range of conditions may lead to a fundamental understanding of the transport phenomena \cite{Qia21,Lin14}. The lack of knowledge of the importance of out-of-plane acoustic phonons seriously limits the efforts to estimate the effects of phonon branches and scattering on the lattice thermal conductivity of graphene.

This study focuses primarily upon the effect of out-of-plane acoustic phonons on the thermal transport properties of graphene. The thermal conductivity at different temperatures and frequencies and in different crystallite sizes is evaluated using a microscopic model that incorporates various phonon modes. The Boltzmann transport equation is solved within the framework of three-phonon interactions without taking into account the four-phonon scattering process. The Umklapp and normal collisions are treated rigorously. The relaxation-time and long-wavelength approximations are avoided for the scattering mechanisms. Calculation results are presented and discussed briefly in relation to theory. This study aims at understanding the importance of out-of-plane acoustic phonons in  thermal transport properties of graphene. Particular emphasis is placed upon the contributions made by different phonon modes to the thermal conductivity under different conditions.

\section{Methods}

Recent studies have demonstrated the quadratic nature of the long wavelength flexural dispersions in low dimensional lattices, irrespective of the lattice type \cite{Kua20}. For example, the flexure branch has quadratic dispersion for the graphene lattice in the long-wavelength region \cite{Kua20}. Over a wide range of the Brillouin zone, the out-of-plane acoustic phonons have quadratic dispersion
\begin{align}
\label{delta-def}
\omega_{ZA}( \vec{q} )=\alpha_{ZA}\vec{q}^{2},
\end{align}
in which $\omega$ denotes the phonon frequency, $\vec q$ denotes the wave vector, and $\alpha _{ZA}$ is a positive constant.

The quadratic dispersion is of importance in the study of  thermal transport of graphene \cite{Mor21,Tah21}. For the out-of-plane acoustic phonons, the density of states is constant
\begin{align}
\label{delta-def}
{D_{ZA}}( \omega  ) = \frac{1}{{4\piup }}\frac{1}{{{\alpha _{ZA}}}}.
\end{align}

The number of the out-of-plane acoustic phonons is given by
\begin{align}
\label{delta-def}
{N_{ZA}}( \omega  )={N_0}( \omega  ){D_{ZA}}( \omega ),
\end{align}
where ${N_0}\left( \omega  \right)$ is the Bose-Einstein distribution function. This function in mode $\lambda \left( {j,\vec q} \right)$ can be written with respect to the Boltzmann constant $k_\text{B}$ as
\begin{align}
\label{delta-def}
{N_{0,\lambda }}=\frac{1}{{{\re^{\frac{{\hbar {\omega _\lambda }}}{{{k_\text{B}}T}}}}-1}},
\end{align}
wherein $j$  is the phonon mode index, $\lambda$ is the phonon mode, $T$ is the temperature, and $\hbar$ is the reduced Planck's constant.

As the frequency tends to 0, the number of the out-of-plane acoustic phonons tends to the reciprocal of frequency and thus the phonon number diverges. Conversely, the acoustic phonons that vibrate in the plane have linear dispersion
\begin{align}
\label{delta-def}
\omega^{'}( {\vec q} ) \approx v^{'}_{g}\vec q,
\end{align}
\begin{align}
\label{delta-def}
D'(\omega)=\frac{1}{2\piup }\frac{\omega }{{v'}_g}^2,
\end{align}
where $v_g$ denotes the phonon group velocity. As the frequency tends to 0, the number of the acoustic phonons that vibrate in the plane approaches a constant. As the frequency tends to 0, the ratio of the above two phonon numbers diverges
\begin{align}
\label{delta-def}
\frac{{{N_{ZA}}\left( \omega  \right)}}{{N'\left( \omega  \right)}}=\frac{1}{2}\frac{{{v'_g}^2}}{{{\alpha _{ZA}}\omega }}.
\end{align}

At the boundary of the Brillouin zone, the ratio of the two phonon numbers is higher than 3. In the frequency range of the out-of-plane acoustic phonons at room temperature determined by the empirical interatomic potential for graphene \cite{Ter88,Lin10b}, the ratio of the two phonon numbers is about 9. When the acoustic phonons that vibrate in the plane are specified as the longitudinal mode, the ratio of the two phonon numbers is higher than 20 and 7, respectively, in the above two cases. Consequently, the out-of-plane acoustic phonons at any temperature are vastly in the majority if the phonons are thermally excited. When the magnitude of wave vector is very small, the group velocity of the out-of-plane acoustic phonons near the boundary of the Brillouin zone exceeds 8000 meters per second, although it may often be negligible \cite{Cah03,Cah14}. This group velocity is more than a half of that of the transverse acoustic phonons and thus it is far from negligible.

One of the most important in the calculation is the selection rule that specifies phonon-phonon scattering in the lattice. The selection rule narrowly limits the possibility of phonon scattering \cite{Lou65,Zak66}. The potential energy of the lattice may be given by
\begin{align}
\label{delta-def}
\Phi \left( { \ldots {{\vec r}_{{l_i}{k_i}}} \ldots } \right)=\sum\limits_{n=2}^\infty  {\frac{1}{{n!}}\sum\limits_{{l_1}{k_1}}^{{l_n}{k_n}} {\sum\limits_{{\alpha _1}}^{{\alpha _n}} {{\Phi _{{\alpha _1} \ldots {\alpha _n}}}\left( {{l_1}{k_1}, \ldots ,{l_n}{k_n}} \right){u_{{\alpha _1}}}\left( {{l_1}{k_1}} \right) \ldots {u_{{\alpha _n}}}\left( {{l_n}{k_n}} \right)} } },
\end{align}
\begin{align}
\label{delta-def}
{\vec r_{lk}}={\vec R_{lk}}+{\vec u_{lk}},
\end{align}
in which $n$ denotes the order of the sequence, $\vec r$ denotes the instantaneous location, $\vec u$ denotes the instantaneous displacement from equilibrium, $\vec R$ denotes the equilibrium position, $k$ denotes the $k$-th atom, and $l$ denotes the $l$-th unit cell of the lattice. Both harmonic and anharmonic terms are contained in the above potential energy expression. More specifically, the second-order series are the harmonic term, and the third-order and higher-order series are the anharmonic term. Both harmonic and anharmonic interatomic force constants are required to calculate the scattering rates and phonon spectrum \cite{Van16,Lee17}. The $n$-th order interatomic force constant can be estimated as
\begin{align}
\label{delta-def}
{\Phi_{{\alpha _1} \ldots {\alpha _n}}}\left( {{l_1}{k_1}, \ldots ,{l_n}{k_n}} \right)=\frac{{{\partial ^n}\Phi }}{{\partial {u_1}\left( {{l_1}{k_1}} \right) \ldots \partial {u_n}\left( {{l_n}{k_n}} \right)}},
\end{align}
in which the partial derivative is evaluated at the equilibrium position of the graphene lattice.

Symmetry considerations impose invariant potentials. A relation between $n$-th order interatomic force constants can be yielded \cite{Lei61,Pat65}:
\begin{align}
\label{delta-def}
\sum\limits_{\alpha^{'}_1, \ldots ,{\alpha^{'}_n}} {{\Phi _{{\alpha^{'}_1}, \ldots ,{\alpha^{'}_n}}}\left( l^{'}_1 k^{'}_1, \ldots , l^{'}_n k^{'}_n \right){\Omega _{{\alpha^{'}_1}}}_{{\alpha _1}} \ldots {\Omega _{{\alpha^{'}_n}}}_{{\alpha _n}}} ={\Phi _{{\alpha _1}, \ldots ,{\alpha _n}}}\left( {{l_1}{k_1}, \ldots ,{l_n}{k_n}} \right),
\end{align}
wherein $\Omega _{\alpha \beta }$ is the matrix corresponding to rotation and reflection \cite{Lei61,Pat65}. This means that the matrix may represent a pure rotation, a pure inversion, or both.

For the lattice, the inversion is a symmetry operation. For a two-dimensional lattice in the $xy$ plane, reflection about the $z$-axis maps the lattice into itself. The mathematical expression for the invariance of the matrix toward reflection is given by
\begin{align}
\label{delta-def}
{\Phi _{{\alpha _1} \ldots {\alpha _n}}}\left( {{l_1}{k_1}, \ldots ,{l_n}{k_n}} \right)={\left( {-1} \right)^m}{\Phi _{{\alpha _1} \ldots {\alpha _n}}}\left( {{l_1}{k_1}, \ldots ,{l_n}{k_n}} \right),
\end{align}
in which $m$ is the total number of the out-of-plane component of the sequence ${\alpha _1} \ldots {\alpha _n}$. If the total number of the out-of-plane component is odd, the above expression becomes
\begin{align}
\label{delta-def}
{\Phi _{{\alpha _1} \ldots {\alpha _n}}}\left( {{l_1}{k_1}, \ldots ,{l_n}{k_n}} \right)=0.
\end{align}

In harmonic approximation, the out-of-plane modes are decoupled from the in-plane modes, since all the mode-mixed terms in the interatomic force constant expression do vanish \cite{Fas07,Mar08}. Additionally, the number of the anharmonic terms can also be reduced by taking into account only the total number of the out-of-plane component that is even. The selection rule specifies that in the complete order of anharmonic phonon-phonon scattering expression, only out-of-plane phonons for which the total number of the out-of-plane component is even are involved due to restrictions governing the anharmonic terms. For example, the following three-phonon interaction processes involving an additional out-of-plane acoustic mode do not occur in the lattice
\begin{align}
ZA+( {LA \Leftrightarrow LA} ),
\end{align}
\begin{align}
ZA+( {LA \Leftrightarrow TA}),
\end{align}
\begin{align}
ZA+( {TA \Leftrightarrow TA} ),
\end{align}
\begin{align}
ZA+( {ZA \Leftrightarrow ZA} ),
\end{align}
\begin{align}
ZA+( {ZA \Leftrightarrow ZO}),
\end{align}
\begin{align}
ZA+( {ZO \Leftrightarrow ZO} ).
\end{align}

Since the intrinsic thermal resistance results from the phonon-phonon scattering in the lattice, the effect of the selection rule on the phonon transport process in the crystal lattice needs to be evaluated. In the lattice potential, only the anharmonic term containing the third-order series is used to calculate the thermal conductivity in order to evaluate the significance of the selection rule and the high density of states in the out-of-plane acoustic mode. Heat is transported by phonons in a graphene ribbon with an infinite length and a finite width along the width direction. The assumed mechanisms of the scattering of lattice waves are Umklapp, normal, and boundary scattering \cite{Alo13,Mic15}. Accordingly, the thermal transport by phonons in the lattice is governed by the anharmonicity of the lattice force and by the boundary of the lattice. The scattering of lattice waves from imperfections in the lattice gives rise to thermal resistance to phonon transport \cite{Fth12,Fth14}. In the present study, however, this scattering mechanism is not taken into account.

The phonon distribution is perturbed by applying a small gradient of temperature to the lattice. The temperature gradient allows phonons to diffuse in the lattice, and phonons scatter inelastically from each other due to the anharmonicity of the interatomic potential \cite{Fah04,Ram16}. The Boltzmann transport equation is solved iteratively within the framework of three-phonon interactions without taking into account the four-phonon scattering process. The Umklapp and normal collisions are treated rigorously in order to avoid relaxation-time approximations. The linearized Boltzmann transport equation is solved for the lattice subjected to a thermal gradient \cite{Bro07,War09}:
\begin{align}
{k_\text{B}}T{v_\lambda }\nabla T\frac{{\partial {N_{0,\lambda }}}}{{\partial T}}&=\sum\limits_{\lambda '\lambda ''} {\Bigl( {W_{\lambda \lambda '\lambda ''}^+\left( {{\Psi _{\lambda ''}}-{\Psi _{\lambda '}}-{\Psi _\lambda }} \right)+\frac{1}{2}W_{\lambda \lambda '\lambda ''}^-\left( {{\Psi _{\lambda ''}}+{\Psi _{\lambda '}}-{\Psi _\lambda }} \right)} \Bigr)} \nonumber\\
& - {N_{0,\lambda }}\left( {{N_{0,\lambda }}+1} \right){\Psi _\lambda }\frac{1}{{\tau _\lambda ^{bs}}},
\end{align}
\begin{align}
\tau _\lambda ^{bs}=\frac{1}{2}\frac{w}{{v_\lambda ^\alpha }},
\end{align}
where $v_\lambda$ is the phonon velocity in mode $\lambda \left( {j,\vec q} \right)$, $\tau$ is the phonon scattering time, $bs$ indicates the boundary scattering, $w$ is the width, $\alpha$ indicates the direction of heat flow, $W$ is three-phonon scattering rate, and $\Psi$ is the nonequilibrium distribution function. The phonon scattering time is determined by taking into account both theoretically ballistic and diffusive transport limits \cite{Min05}. The phonon scattering time is directly proportional to the nonequilibrium distribution function \cite{Bro05}. The three-phonon interactions can be represented by the collision term \cite{Omi96,Spa02}. More specifically, the anharmonic three-phonon scattering process is described by the above linearized equation using the nonequilibrium distribution function. On the right-hand side of the linearized equation, the summation term describes the three-phonon scattering processes, and the negative term describes the boundary scattering process phenomenologically with the phonon scattering time.

In the scattering process, phonon energy is conserved. Within the framework of three-phonon interactions, the law of conservation of energy states as follows \cite{Bro05}:
\begin{align}
{\omega _j}\left( {\vec q} \right)\pm{\omega _{j'}}\left( {\vec q'} \right)={\omega _{j''}}( {\vec q''} ).
\end{align}

In addition to conservation of energy, crystal momentum is also conserved in the scattering process. Within the framework of three-phonon interactions, the law of conservation of quasi-momentum can be written as follows \cite{Bro05}:
\begin{align}
\vec q\pm\vec q'=\vec q''+\vec g,
\end{align}
where $\vec g$ denotes a reciprocal-lattice vector. This vector is non-zero for Umklapp scattering. Additionally, this vector is zero for normal scattering, since the total crystal momentum is conserved in the normal processes. The upper and the lower signs refer to the following processes
\begin{align}
\lambda \left( {\vec q} \right)+\lambda \left( {\vec q'} \right) \to \lambda ( {\vec q''} ),
\end{align}
\begin{align}
\lambda \left( {\vec q} \right) \to \lambda \left( {\vec q'} \right)+\lambda ( {\vec q''} ).
\end{align}

The three-phonon scattering rate can be determined by applying Fermi's golden rule
\begin{align}
W_{\lambda \lambda '\lambda ''}^\pm=\frac{{\hbar \piup }}{{4N'}}\frac{1}{{{\omega _\lambda }{\omega _{\lambda '}}{\omega _{\lambda ''}}}}\left( {{N_{0,\lambda }}+1} \right)\left( {{N_{0,\lambda '}}+\frac{1}{2}\pm\frac{1}{2}} \right){N_{0,\lambda ''}}{\left( {\Phi _{\lambda \lambda '\lambda ''}^\pm} \right)^2}\delta \left( {{\omega _\lambda }\pm{\omega _{\lambda ''}}-{\omega _{\lambda '}}} \right),
\end{align}
in which the delta function, $\delta \left( {{\omega _\lambda }\pm{\omega _{\lambda ''}}-{\omega _{\lambda '}}} \right)$, ensures conservation of energy and $N'$ denotes the number of unit cells of the lattice.

The scattering intensity is indicated by the three-phonon scattering matrix elements
\begin{align}
\Phi _{\lambda \lambda '\lambda ''}^\pm=\sum\limits_k {\sum\limits_{l'k'} {\sum\limits_{l''k''} {\sum\limits_{\alpha \beta \gamma } {{\Phi _{\alpha \beta \gamma }}\left( {0k,l'k',l''k''} \right){\re^{\ri\vec q' \cdot {{\vec R}_{l'}}}} \cdot {\re^{\ri\vec q'' \cdot {{\vec R}_{l''}}}} \cdot \left( {e_{\alpha k}^\lambda e_{\beta k'}^{\pm\lambda '}e_{\gamma k''}^{-\lambda ''}} \right) \cdot \Xi } } } }\,,
\end{align}
\begin{align}
\Xi ={\left( {{M_k}} \right)^{-\frac{1}{2}}} \cdot {\left( {{M_{k'}}} \right)^{-\frac{1}{2}}} \cdot {\left( {{M_{k''}}} \right)^{-\frac{1}{2}}},
\end{align}
\begin{align}
\lambda  \to -\lambda  \Rightarrow \vec q \to -\vec q,
\end{align}
wherein $\Xi$ denotes a composite parameter,  $\alpha$,  $\beta$, $\gamma$ are Cartesian components, $M$ denotes the atom mass, $\vec R$ denotes a lattice vector, $e_{\alpha k}^\lambda$ denote phonon eigenvectors, and ${\Phi _{\alpha \beta \gamma }}\left( {0k,l'k',l''k''} \right)$ denote the real-space anharmonic interatomic force constants.

The Boltzmann transport equation is solved using an iterative method \cite{Bro07,War09} so that the nonequilibrium distribution function can be determined. The lattice thermal conductivity of graphene is given by
\begin{align}
k=\sum\nolimits_j {{k_j}},
\end{align}
\begin{align}
{k_j}=\frac{{{k_\text{B}}}}{{4{\piup ^2}}}\frac{1}{{\delta '}}{\int {\left( {\frac{{\hbar \omega \left( {j,\vec q} \right)}}{{{k_\text{B}}T}}} \right)}^2}\frac{1}{{{\re^{\frac{{\hbar \omega \left( {j,\vec q} \right)}}{{{k_\text{B}}T}}}}-1}}\left( {\frac{1}{{{\re^{\frac{{\hbar \omega \left( {j,\vec q} \right)}}{{{k_\text{B}}T}}}}-1}}+1} \right){v^2}\left( {j,\vec q} \right)\tau \left( {j,\vec q} \right)d\vec q,
\end{align}
wherein $\tau \left( {j,\vec q} \right)$ is the scattering time in mode $\lambda \left( {j,\vec q} \right)$, and $\delta '$, 0.335 nm, is the thickness of the lattice.

The anharmonic and harmonic interatomic force constants need to be determined in order to calculate the phonon eigenvectors, phonon frequencies, and three-phonon scattering rates. These interatomic force constants are calculated using an optimized Tersoff empirical interatomic potential in order to better fit the acoustic phonon velocities and frequencies. To validate the model described above, independent calculations of the anharmonic and harmonic interatomic force constants and the thermal conductivity are also performed using a real space approach within the generalized gradient approximation and using a projector-augmented wave pseudopotential implemented with the Vienna Ab initio Simulation Package. Good agreement is achieved between the results obtained from the optimized Tersoff empirical interatomic potential and from first principles.

\section{Results and discussion}

\begin{figure}[!t]
\centerline{\includegraphics[width=0.55\textwidth]{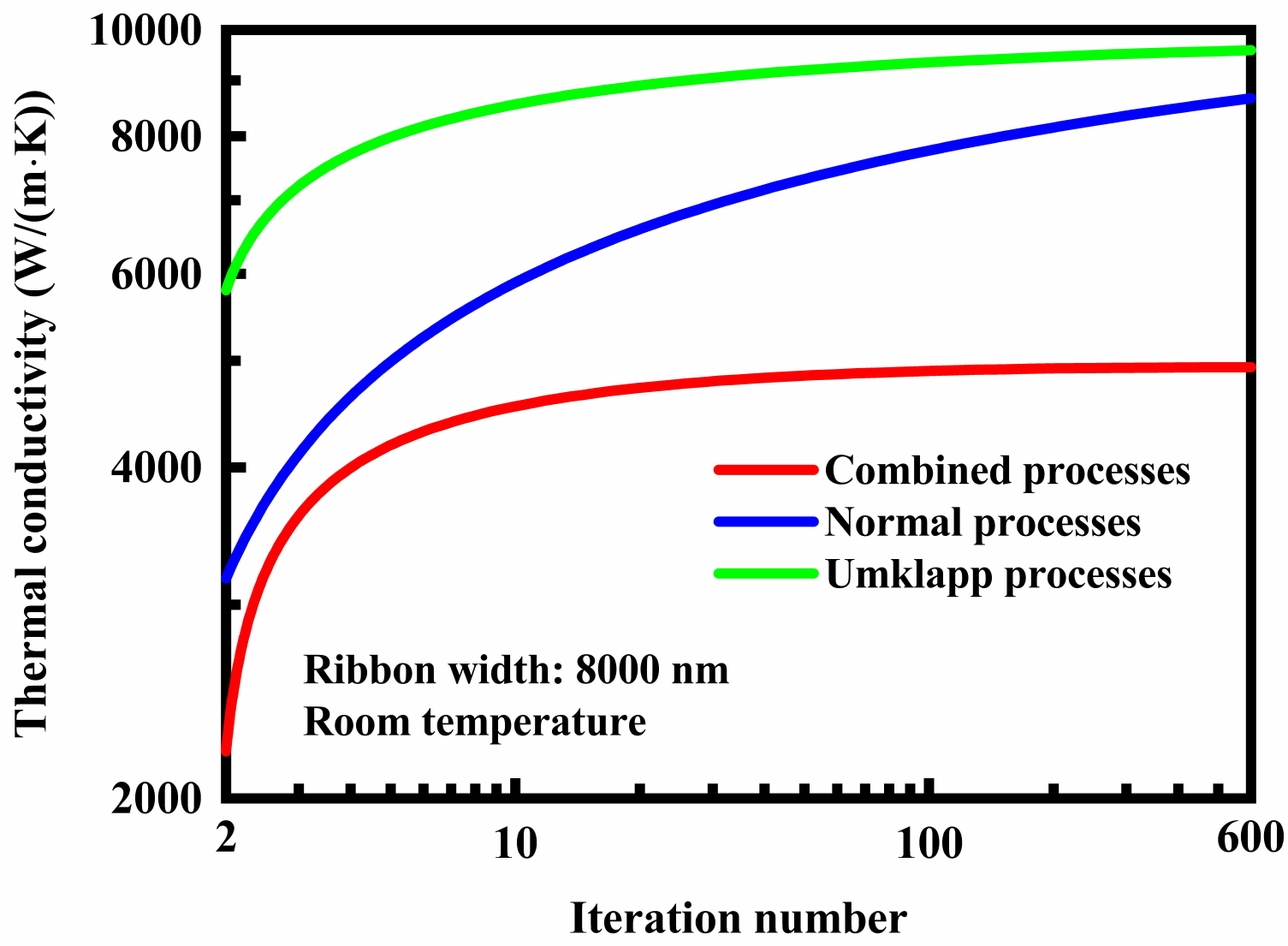}}
\caption{(Colour online) Effect of iteration number on the thermal conductivity of the graphene ribbon at room temperature with taking into account different phonon scattering processes. The ribbon is 8000 nm in width, and calculations are made by assuming Umklapp processes, normal processes, and a combination of them.} \label{fig-smp1}
\end{figure}

The effect of iteration number on the thermal conductivity at room temperature is illustrated in figure~\ref{fig-smp1} for the graphene ribbon with taking into account different phonon scattering processes. The ribbon is 8000 nm in width, and calculations are made by assuming Umklapp processes, normal processes, and a combination of them. In the combined case, the full converged solution is 4980 W/(m$\cdot$K) for the thermal conductivity. Additionally, the thermal conductivity converges very rapidly. By contrast, the thermal conductivity does not converge in the specified iteration number range, if only normal processes are taken into account. Additionally, the thermal conductivity increases rapidly with the iteration number. Furthermore, phonon transport may approach the ballistic limit, since normal processes do not contribute to the thermal resistivity. This is because normal scattering merely re-distributes the energy into different phonon modes without changing its total flow. The thermal conductivity does not converge very rapidly, if only Umklapp processes are taken into account. The thermal conductivity in this case is much higher than that in the combined case in which both Umklapp and normal processes are taken into account. The interactions between Umklapp and normal processes are of importance in the study of the phonon transport properties. Whereas normal processes make no contribution to the thermal resistance, the wave vector changes with normal scattering so that any extra Umklapp processes will provide additional phonon scattering. Both Umklapp and normal processes must be taken into account to accurately predict the phonon transport properties of graphene.

\begin{figure}[!t]
\centerline{\includegraphics[width=0.55\textwidth]{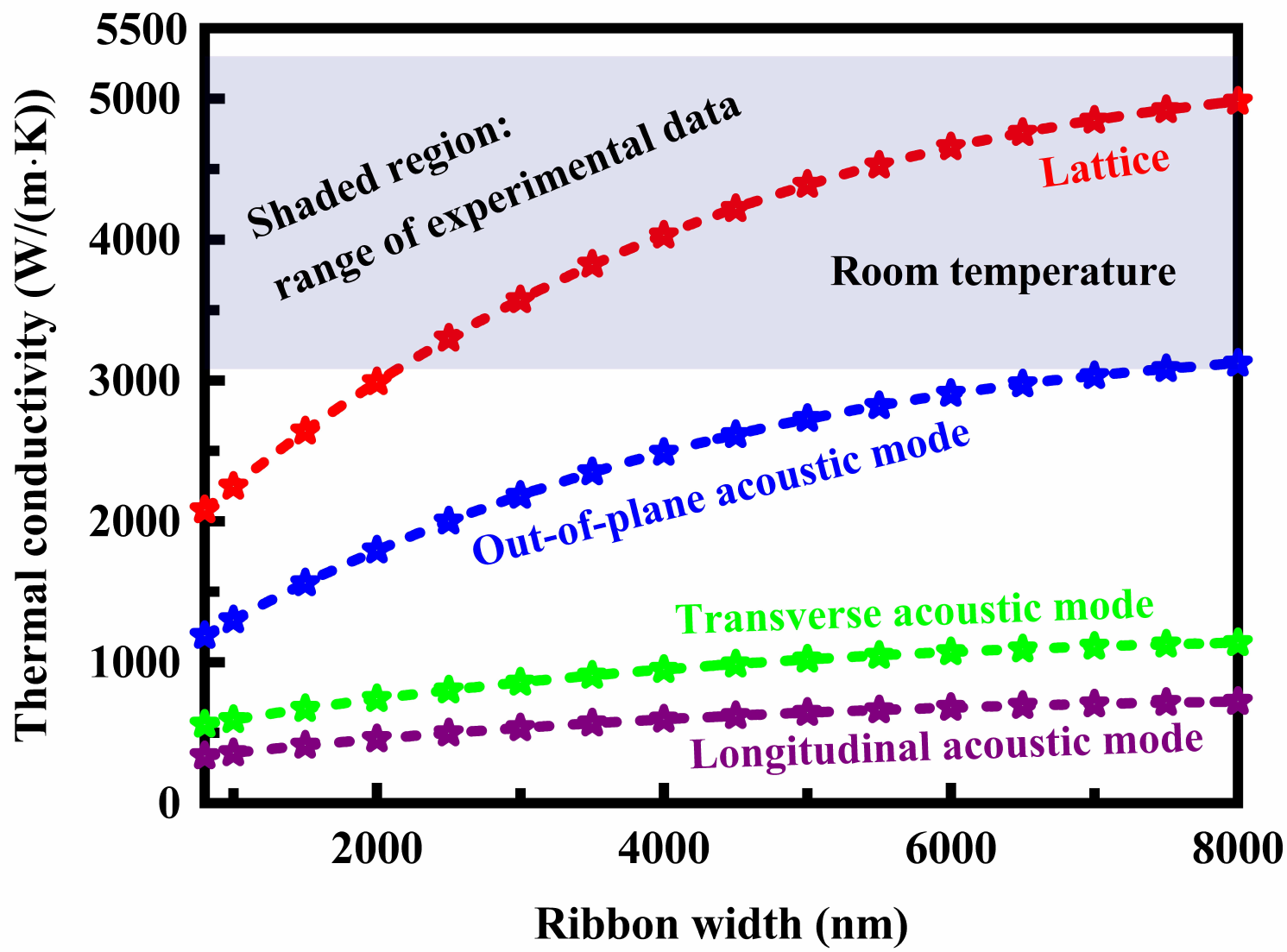}}
\caption{(Colour online) Thermal conductivity of the graphene ribbon with different widths at room temperature. The thermal conductivity in different modes is also presented. The range of experimental data is indicated by the shaded region.} \label{fig-smp2}
\end{figure}

The thermal conductivity results at room temperature are presented in figure~\ref{fig-smp2} with different widths. The phonon thermal conductivity results in different modes are also presented. The ribbon width varies from 800 to 8000 nm. The range of experimental data \cite{Bal08,Gho08} is indicated by the shaded region. Reasonably good agreement with the experimental data is obtained. The out-of-plane acoustic phonons make a great contribution to the thermal conductivity, and the contributions made by both the longitudinal and transverse acoustic phonons are relatively small. In addition, the optical phonons are incapable of contributing to the thermal conductivity. This is because all the three-phonon interaction processes involving one or three out-of-plane phonon do not occur in the lattice, as specified by the selection rule. The selection rule governs the likelihood that a three-phonon interaction process will occur in the lattice. The selection rule also specifies the forbidden three-phonon interaction processes that the total number of the out-of-plane phonons is one or three. Accordingly, the phase space for three-phonon Umklapp and normal scattering is dramatically reduced due to restrictions governing the anharmonic terms. Little space is left for three-phonon scattering involving the out-of-plane acoustic mode to the thermal resistance.

\begin{figure}[!t]
\centerline{\includegraphics[width=0.55\textwidth]{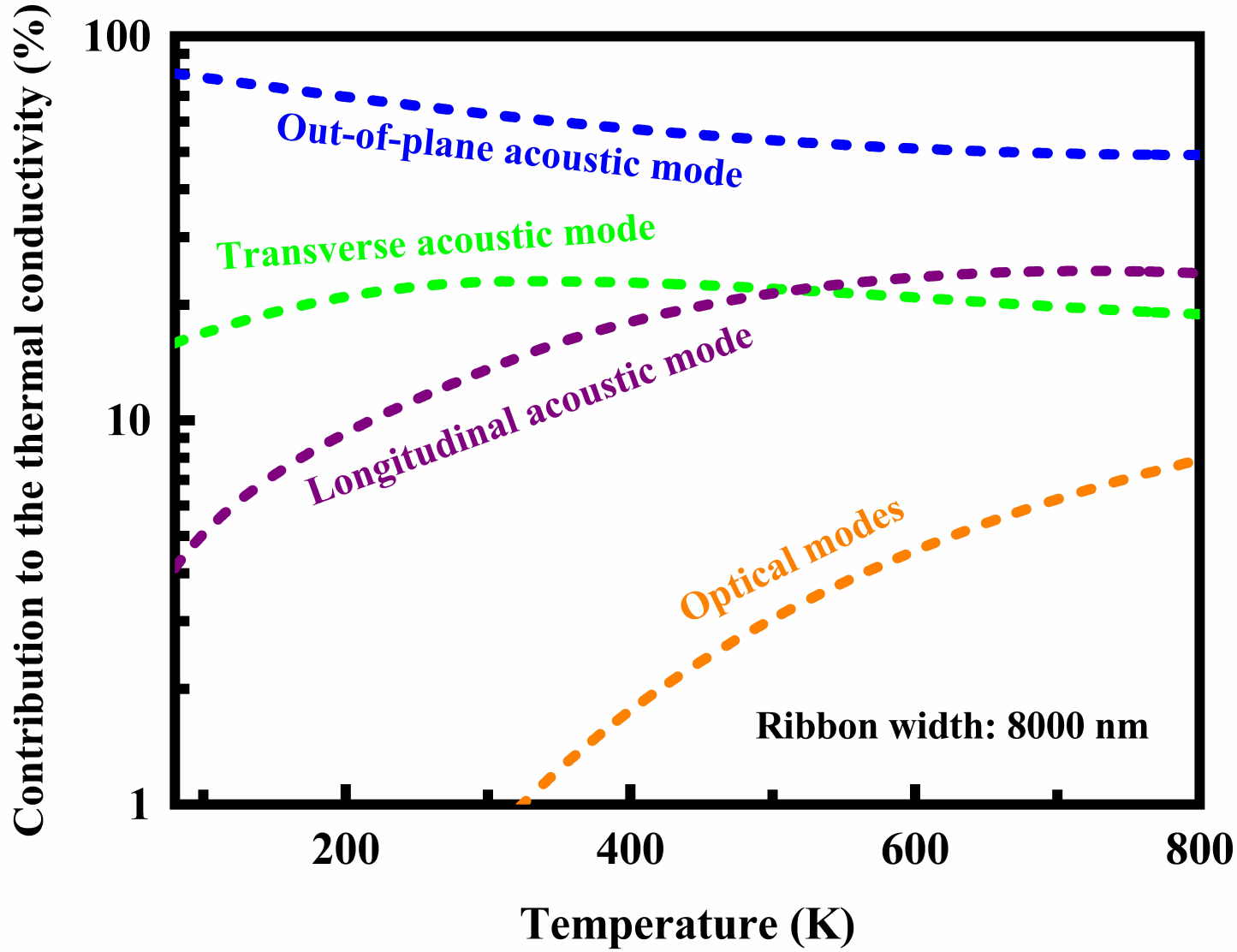}}
\caption{(Colour online) Phonon mode contributions to the thermal conductivity of the graphene ribbon at different temperatures.} 
\label{fig-smp3}
\end{figure}

The contributions made by different phonon modes to the thermal conductivity are illustrated in figure~\ref{fig-smp3} at different temperatures. The ribbon is 8000 nm in width. The phonon mode contributions are scaled by the thermal conductivity. At lower temperatures, the out-of-plane acoustic phonons make a significant contribution to the thermal conductivity, and the longitudinal and transverse acoustic phonons make small contributions. The optical phonons make no contribution to the thermal conductivity. The thermal conductivity in each mode changes with respect to temperature. For example, as the temperature increases, the thermal conductivity in the out-of-plane acoustic mode decreases. The contribution made by the optical phonons cannot be neglected especially at higher temperatures. However, the contribution made by the out-of-plane acoustic phonons is substantially greater than or equal to the total contribution made by the other phonons. At higher temperatures, the contribution made by phonons in each acoustic mode remains almost unchanged. This is because under such conditions, the thermal conductivity in each acoustic mode has a substantial inverse dependence on temperature. There is a complex relationship between the thermal conductivity and temperature. At lower temperatures, the thermal conductivity is proportional to temperature squared \cite{Nik09a,Nik09b}. At moderate and higher temperatures, the effect of Umklapp scattering on the thermal properties of graphene becomes pronounced \cite{Nik09a,Nik09b}. Accordingly, the thermal conductivity is inversely proportional to temperature due to the increased strength of the Umklapp scattering processes.

\begin{figure}[!t]
	\centerline{\includegraphics[width=0.55\textwidth]{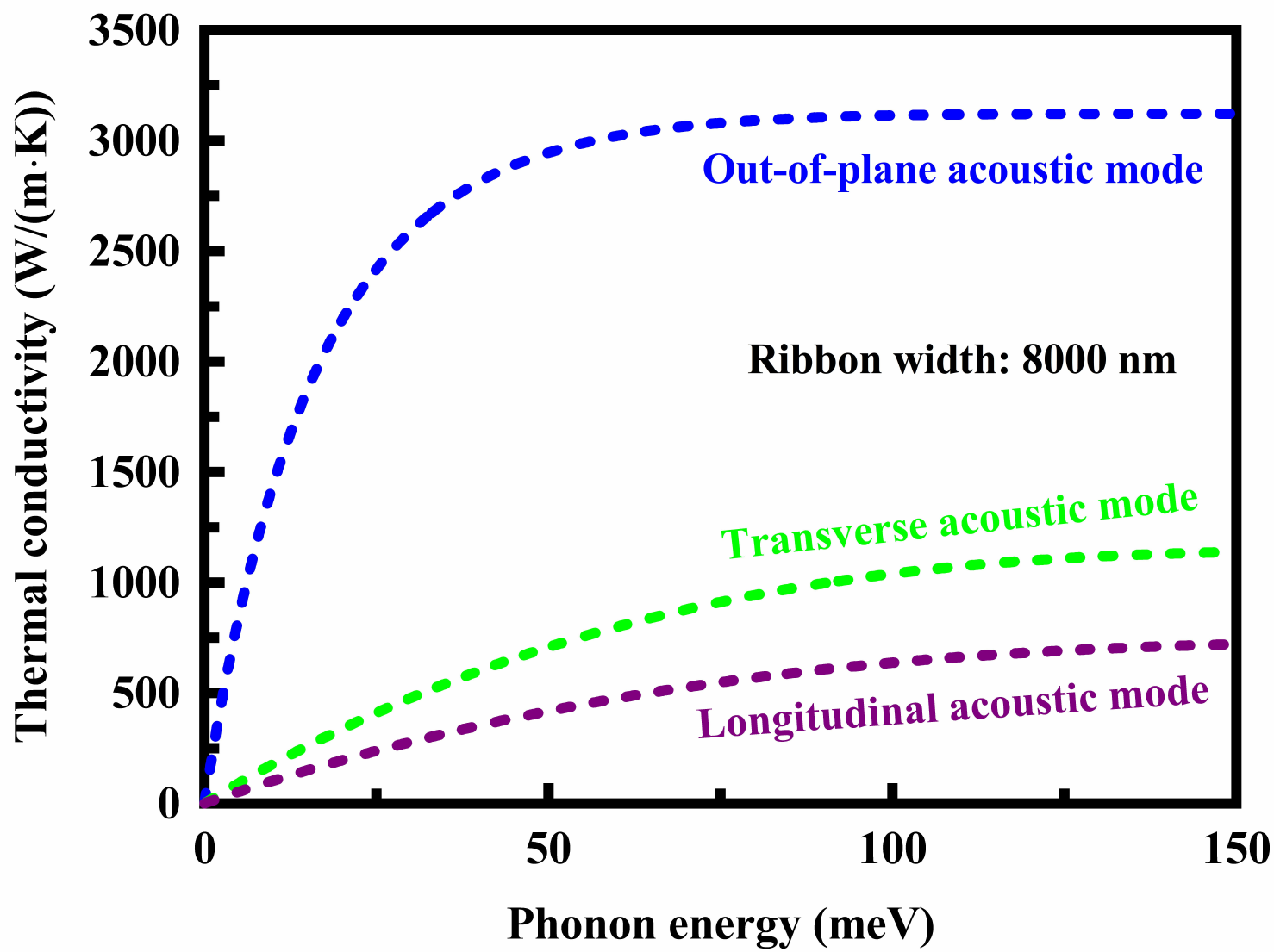}}
	\caption{(Colour online) Thermal conductivity in each mode for the graphene ribbon at room temperature under different phonon energy conditions. The thermal conductivity in each mode is calculated by a mere integration over the entire phonon frequency spectrum.} 
	\label{fig-smp4}
\end{figure}

Knowledge of the effect of phonon frequency on the thermal conductivity is essential to understand the thermal transport process in the graphene lattice \cite{Pop10,Ram20}. This study  explicitly concerns  the calculation of the thermal conductivity at different phonon frequencies. The contributions made by different phonon modes to the thermal conductivity at room temperature are illustrated in figure~\ref{fig-smp4} under different phonon energy conditions. The ribbon is 8000 nm in width, and the phonon frequency is expressed in terms of phonon energy. The thermal conductivity is calculated at room temperature and thus the contribution made by the optical phonons can be neglected. The thermal conductivity in each mode is calculated by a mere integration over the entire phonon frequency spectrum. The limits of integration are zero and the specified phonon frequency. In the Brillouin zone, the out-of-plane acoustic phonons make a great contribution to the thermal conductivity over the full phonon frequency range. In addition, the total contribution made by the other phonons is relatively small in the frequency spectrum, including both the longitudinal and transverse acoustic modes. As the frequency increases, the thermal conductivity in each mode increases. At higher frequencies, the thermal conductivity in the out-of-plane acoustic mode remains constant. This is because the out-of-plane acoustic phonons are enriched in the lower frequency range. Accordingly, the intensity of normal scattering is much higher than that of Umklapp scattering.

\begin{figure}[!t]
\centerline{\includegraphics[width=0.55\textwidth]{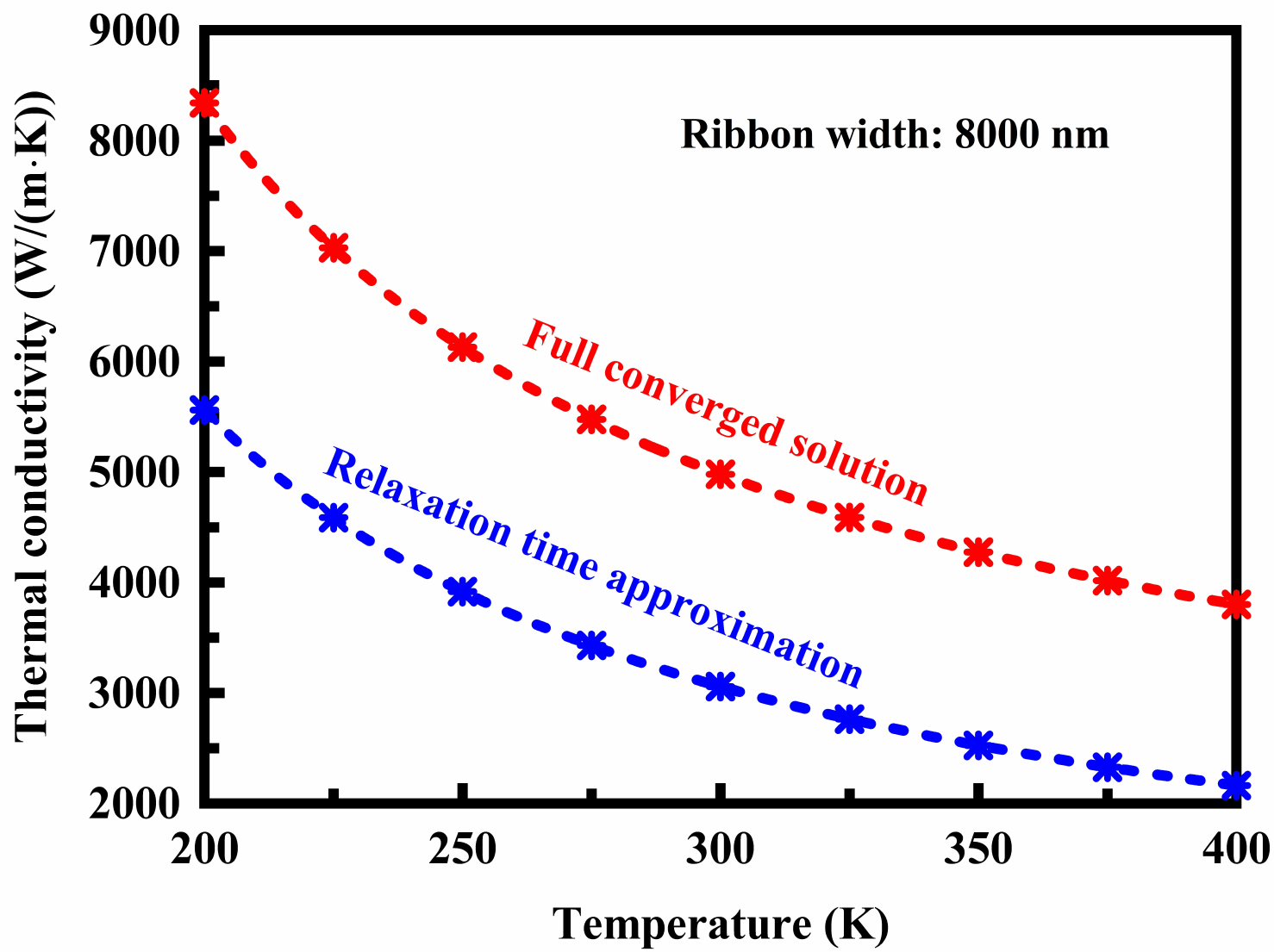}}
\caption{(Colour online) Thermal conductivity of the graphene ribbon at different temperatures obtained from the full converged solution and obtained with the use of the relaxation time approximation.} 
\label{fig-smp5}
\end{figure}

The thermal conductivity results at room temperature are presented in figure~\ref{fig-smp5} from the full converged solution and with the use of the relaxation time approximation. For the graphene lattice, the thermal conductivity is obtained from the full converged solution to the Boltzmann transport equation. Additionally, the solution to the thermal transport problem is also solved with the use of the approximation. The thermal conductivity obtained from the full converged solution is much higher than that obtained with the use of the approximation. The failure of the relaxation time approximation in describing the phonon transport process in the lattice is discussed here. This method treats three-phonon normal scattering as independent, resistive processes. Accordingly, physically incorrect results will be produced with this approximation. For example, treatments with only three-phonon normal processes make the thermal conductivity divergent without taking into account both Umklapp and boundary scattering. For most semiconductors at room temperature, corrections that are needed to be made to the solution are often small with the use of the relaxation time approximation \cite{Cal59,War10} due to the strong scattering intensity. For the graphene lattice, however, the relaxation time approximation is by no means an accurate simplification, as discussed below.

\begin{figure}[!t]
	\centerline{\includegraphics[width=0.55\textwidth]{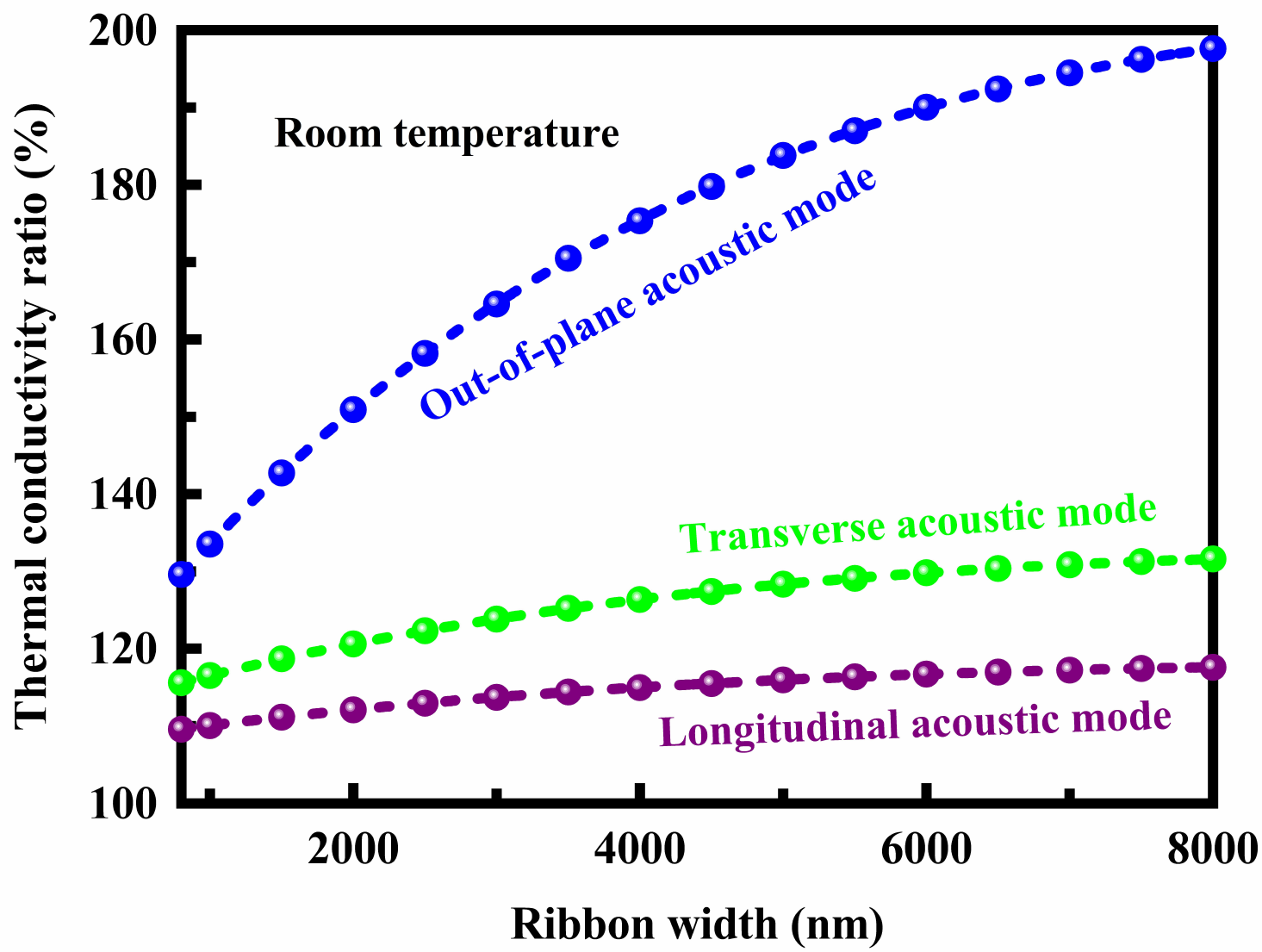}}
	\caption{(Colour online) Thermal conductivity ratio in different phonon modes at room temperature for the graphene ribbon with different widths. The thermal conductivity ratio is defined as the ratio of the thermal conductivity obtained from the full converged solution to that obtained with the use of the relaxation time approximation.} 
	\label{fig-smp6}
\end{figure}

The thermal conductivity ratio results in different phonon modes at room temperature are presented in figure~\ref{fig-smp6} for the graphene ribbon with different widths. The thermal conductivity ratio is defined as the ratio of the thermal conductivity obtained from the full converged solution to that obtained with the use of the relaxation time approximation. The thermal conductivity ratio depends upon the ribbon width. In the out-of-plane acoustic mode, there is a significant difference in thermal conductivity between the iteration and approximate solutions. More specifically, the thermal conductivity obtained from the full converged solution is much higher than that obtained with the use of the relaxation time approximation, with up to about a twofold increase. In the longitudinal or transverse acoustic mode, the difference is relatively small but there is still an increase in thermal conductivity. This is because an increase in the density of out-of-plane acoustic phonons will lead to an increase in the density of longitudinal or transverse acoustic phonons due to the dynamic equilibrium in the three-phonon scattering process. The failure of the relaxation time approximation is caused by the elastic scattering assumption. The assumption is not justified since anharmonic phonon-phonon scattering is an inelastic process.

\begin{figure}[!t]
	\centerline{\includegraphics[width=0.55\textwidth]{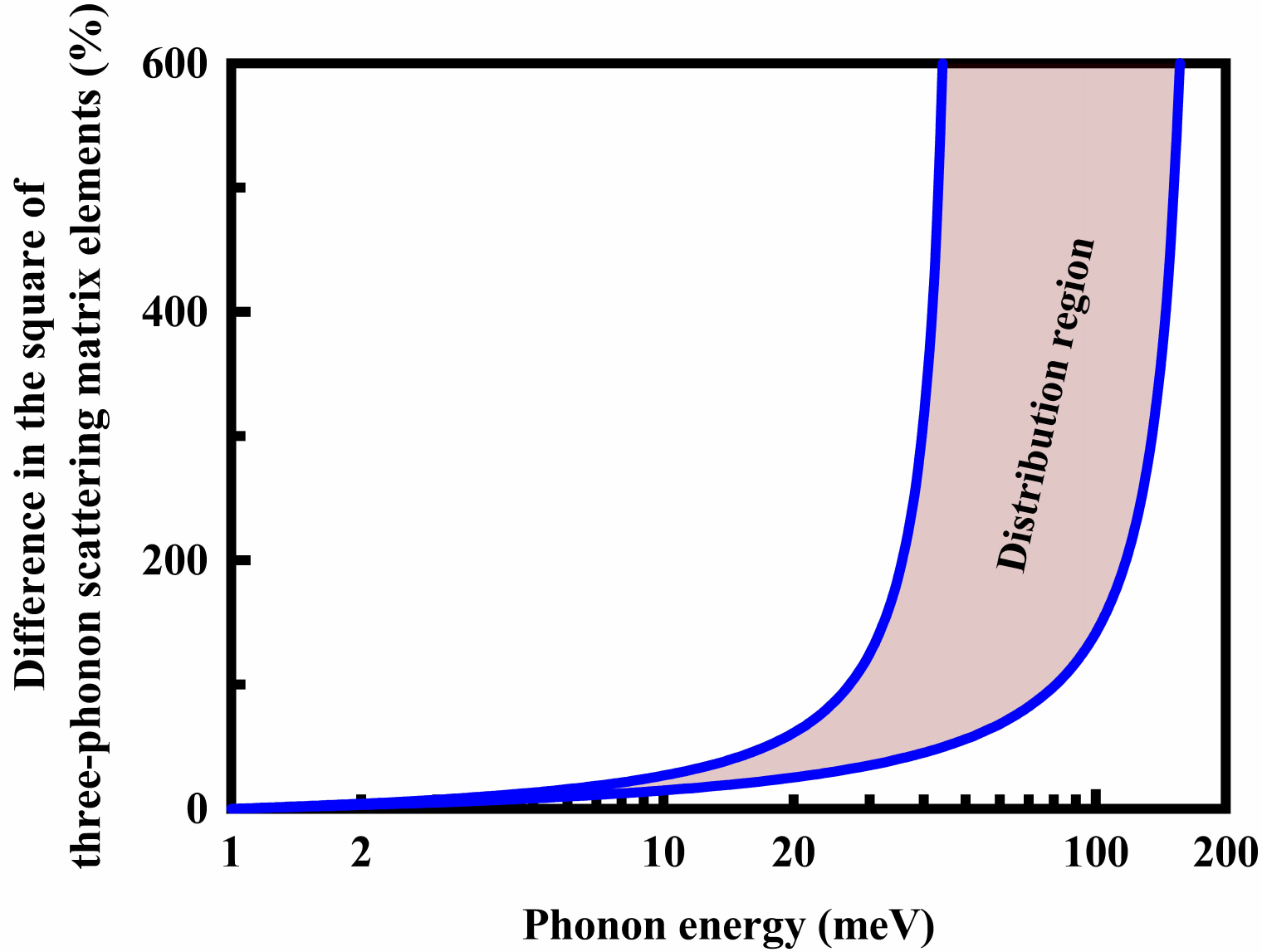}}
	\caption{(Colour online) Difference in the square of three-phonon scattering matrix elements between the long wavelength approximation and the full converged solution for the graphene ribbon under different phonon energy conditions.} 
	\label{fig-smp7}
\end{figure}

Three-phonon scattering rates can also be determined with the use of the long wavelength approximation \cite{Kle58,Sin11}. This approximation is made that only the acoustic branches of the phonon spectrum are taken into account and the lattice is treated as an elastic continuum. The magnitudes of all the phonon wave vectors involved in each scattering process are much smaller than the reciprocal of the lattice spacing. However, the magnitudes of phonon wave vectors must be sufficiently large if three-phonon Umklapp processes occur. These processes make great contribution to the thermal resistance, but the long wavelength approximation fails to make accurate predictions. The difference in the square of three-phonon scattering matrix elements between the long wavelength approximation and the full converged solution is illustrated in figure~\ref{fig-smp7} for the graphene ribbon under different phonon energy conditions. The phonon frequency is expressed in terms of phonon energy. The long wavelength approximation is only justified in normal processes in which the phonon frequencies are low. At higher phonon frequencies, the long wavelength approximation fails to make accurate predictions. Consequently, the accuracy of this method depends heavily upon the phonon frequency. The long wavelength approximation has the great disadvantage of extreme simplification without taking into account normal scattering \cite{Kon09,Bal76}. This approximation is made based upon the fact that Umklapp scattering contributes strongly to the thermal resistance but normal scattering has no contribution. However, the wave vector changes with normal scattering so that any extra Umklapp processes will provide additional phonon scattering. For the graphene lattice, the scattering matrix elements are not routinely incorporated into the long wavelength approximation. The matrix elements do not vanish in various cases in which the conservation constraints are satisfied and the selection rule for the interaction between phonons must still pertain.

\section{Conclusions}

The lattice thermal conductivity of graphene at different temperatures and frequencies and in different crystallite sizes is evaluated within the framework of a microscopic model that incorporates both acoustic and optical modes with phonon dispersion relations. The linearized phonon-Boltzmann transport equation is solved iteratively within the framework of three-phonon interactions without taking into account the four-phonon scattering process. The major conclusions are summarized as follows:

\begin{itemize}
\item
The full converged solution is 4980 W/(m$\cdot$K) for the thermal conductivity. Good agreement with the experimental data is obtained. The interactions between Umklapp and normal processes are of importance in the study of the phonon transport properties. Both Umklapp and normal processes must be taken into account to predict the phonon transport properties accurately.

\item
The out-of-plane acoustic phonons contribute greatly to the thermal conductivity and the longitudinal and transverse acoustic phonons make small contributions over a wide range of temperatures and phonon frequencies. The contribution made by the optical phonons cannot be neglected, especially at higher temperatures.

\item
The out-of-plane acoustic phonons dominate the thermal conductivity due to their high density of states and restrictions governing the anharmonic phonon scattering. The selection rule severely restricts the phase space for out-of-plane phonon scattering due to reflection symmetry. Little space is left for out-of-plane phonon scattering to the thermal resistance.

\item
Both relaxation time and long wavelength approximations fail to make accurate predictions about the phonon transport properties. The relaxation time approximation is by no means an accurate simplification. Additionally, the long wavelength approximation is only justified in normal processes in which the phonon frequencies are low.
\end{itemize}

%% Type in your references using {thebibliography} environment 
%% or create them from your bibtex database using cmpj.bst style.

\bibliographystyle{cmpj}
\bibliography{cmpjxampl}

\ukrainianpart

\title{Вплив позаплощинних акустичних фононів на властивості теплопровідності у графені}
\author{Дж. Чен, Ю. Ліу} 

\address{
	Відділення енергетики, Механіко-енергетичного факультет, Хенанський політехнічний університет, Цзяоцзо, Хенань, 454000, КНР
}

\makeukrtitle

\begin{abstract}
	Теплопровідність ґратки графену визначається з використанням мікроскопічної моделі, яка враховує дискретну природу даної ґратки та дисперсійне співвідношення для фононів в межах зони Бріллюена. Рівняння переносу Больцмана розв'язується ітераційно в рамках трифононних взаємодій без врахування чотирифононного розсіювання. Процеси перекиду (Umklapp) та нормальні зіткнення  враховуються точно, уникаючи таким чином апроксимацій для часу релаксації та довгохвильового наближення. Обговорюються причини неспроможності цих наближень передбачити властивості теплопереносу. Визначення теплопровідності здійснюється при різних температурах і частотах та при різних розмірах кристалітів. Отримано прийнятно добре узгодження з експериментальними даними. Обчислення виявили вирішальну роль позаплощинних акустичних фононів для визначення теплопровідності. Внесок цих фононів є значним, тоді як внесок поздовжних і поперечних акустичних фононів є малим в широкому діапазоні температур і частот. Позаплощинні акустичні фонони домінують у формуванні теплопровідності завдяки їх високій густині станів та обмеженням, що регулюють ангармонійне розсіювання фононів. Правило відбору суттєво обмежує фазовий простір для розсіювання позаплощинних фононів через симетрію відбивання. Внеском оптичних фононів не можна нехтувати при вищих температурах. Для точного передбачення властивостей фононного переносу необхідно враховувати як процеси з перекиданням хвильового вектора, так і нормальні процеси.
	\keywords теплові властивості, теплопровідність, фонони, теорія розсіювання, явища переносу, графен
\end{abstract}

  \end{document}